\documentclass{JINST}

\title{Low background detector with enriched $^{116}$CdWO$_4$ crystal
scintillators to search for double $\beta$ decay of $^{116}$Cd}

\author{A.S.~Barabash$^a$,
P.~Belli$^b$,
R.~Bernabei$^{b,c,}$\thanks{Corresponding author.},
R.S.~Boiko$^d$,
F.~Cappella$^{e,f}$,
V.~Caracciolo$^{g,h}$,
D.M.~Chernyak$^d$,
R.~Cerulli$^g$,
F.A.~Danevich$^d$,
M.L.~Di~Vacri$^g$,
A.E.~Dossovitskiy$^i$,
E.N.~Galashov$^j$,
A.~Incicchitti$^{e,f}$,
V.V.~Kobychev$^d$,
S.I.~Konovalov$^a$,
G.P.~Kovtun$^k$,
V.M.~Kudovbenko$^d$,
M.~Laubenstein$^g$,
A.L.~Mikhlin$^i$,
S.~Nisi$^g$,
D.V.~Poda$^{g,d}$,
R.B.~Podviyanuk$^d$,
O.G.~Polischuk$^d$,
A.P.~Shcherban$^k$,
V.N.~Shlegel$^j$,
D.A.~Solopikhin$^k$,
Yu.G.~Stenin$^j$,
V.I.~Tretyak$^d$,
V.I.~Umatov$^a$,
Ya.V.~Vasiliev$^j,$
V.D.~Virich$^k$\\
\llap{$^a$}Institute of Theoretical and Experimental Physics, 117259 Moscow, Russia\\
\llap{$^b$}INFN sezione Roma ``Tor Vergata'', I-00133 Rome, Italy\\
\llap{$^c$}Dipartimento di Fisica, Universit\`a di Roma ``Tor Vergata'', I-00133, Rome, Italy\\
\llap{$^d$}Institute for Nuclear Research, MSP 03680 Kyiv, Ukraine\\
\llap{$^e$}INFN sezione Roma ``La Sapienza'', I-00185 Rome, Italy\\
\llap{$^f$}Dipartimento di Fisica, Universit\`a di Roma ``La Sapienza'', I-00185 Rome, Italy\\
\llap{$^g$}INFN, Laboratori Nazionali del Gran Sasso, I-67100 Assergi (AQ), Italy\\
\llap{$^h$}Dipartimento di Fisica, Universit\`a dell'Aquila, I-67100 L'Aquila, Italy\\
\llap{$^i$}Joint Stock Company NeoChem, 117647 Moscow, Russia\\
\llap{$^j$}Nikolaev Institute of Inorganic Chemistry, 630090 Novosibirsk, Russia\\
\llap{$^k$}National Science Center ``Kharkiv Institute of Physics and Technology'', 61108 Kharkiv,
Ukraine\\
  E-mail: \email{rita.bernabei@roma2.infn.it}}

\abstract{A cadmium tungstate crystal boule enriched in $^{116}$Cd
to 82\% with mass of 1868 g was grown by the low-thermal-gradient
Czochralski technique. The isotopic composition of cadmium and the
trace contamination of the crystal were estimated by High
Resolution Inductively Coupled Plasma Mass-Spectrometry. The
crystal scintillators produced from the boule were subjected to
characterization that included measurements of transmittance and
energy resolution. A low background scintillation detector with
two $^{116}$CdWO$_4$ crystal scintillators (586 g and 589 g) was
developed. The detector was running over 1727 h deep underground
at the Gran Sasso National Laboratories of the INFN (Italy), which
allowed to estimate the radioactive contamination of the enriched
crystal scintillators. The radiopurity of a third $^{116}$CdWO$_4$
sample (326 g) was tested with the help of ultra-low background
high purity germanium $\gamma$ detector. Monte Carlo simulations
of double $\beta$ processes in $^{116}$Cd were used to estimate
the sensitivity of an experiment to search for double $\beta$
decay of $^{116}$Cd.}


\begin{document}

\section{Introduction}

The neutrinoless mode of the double beta decay ($0\nu2\beta$) is a
powerful tool to investigate properties of neutrino and weak
interactions. Even negative results of double $\beta$ decay
experiments provide important information on the absolute scale
and hierarchy of the Majorana neutrino mass, lepton number
conservation, right-handed admixtures in weak interaction,
existence of majorons, and other effects beyond the Standard Model
\cite{DBD}.

The cadmium 116 is one of the most prospective isotopes to search
for $0\nu2\beta$ decay thanks to the high energy of the decay
$Q_{\beta\beta}=2809\pm4$ keV \cite{Aud03}, the relatively large
isotopic abundance $7.49\pm0.18$~\% \cite{Ber11}, the promising
theoretical predictions \cite{Kor07,Sim08,Iac11} and the existence
of cadmium tungstate (CdWO$_4$) crystal scintillators allowing to
realize a calorimetric (``source$~=~$detector'') experiment.

The most sensitive $2\beta$ experiment to search for double beta
decay of $^{116}$Cd was performed in the Solotvina Underground
Laboratory with the help of cadmium tungstate crystal
scintillators enriched in $^{116}$Cd ($^{116}$CdWO$_4$, see
\cite{Dan03a} and references therein). The two neutrino mode of
$2\beta$ decay of $^{116}$Cd was observed with the half-life
$T_{1/2}=(2.9^{+0.4}_{-0.3})\times10^{19}$~yr, and the half-life
limit on $0\nu2\beta$ decay of $^{116}$Cd was set as $T_{1/2}\geq
1.7\times10^{23}$~yr at 90\% C.L. It corresponds to an upper bound
on the effective Majorana neutrino mass $\langle
m_{\nu}\rangle\leq1.7$ eV \cite{Dan03a}. Searches for double
$\beta$ processes in $^{106}$Cd, $^{108}$Cd and $^{114}$Cd were
realized by using low background CdWO$_4$ crystal scintillators
produced from cadmium of the natural composition
\cite{Dan96a,Bel08}. Recently, a cadmium tungstate crystal
scintillator enriched in $^{106}$Cd was developed \cite{Bel10a},
and an experiment to search for double beta processes in
$^{106}$Cd with the help of this detector is in progress in the
Gran Sasso Underground Laboratories \cite{Bel10bc}. In addition,
cadmium tungstate crystal scintillators were also successfully
applied to study the fourth-forbidden $\beta$ decay of $^{113}$Cd
\cite{Dan96b,Bel07} and to detect, for the first time, the
$\alpha$ decay of $^{180}$W with the half-life
$T_{1/2}=(1.1^{+0.9}_{-0.5})\times10^{18}$~yr \cite{Dan03b}.

High sensitivity double $\beta$ experiments require detectors with
maximal concentration of the studied isotope, high detection
efficiency to double $\beta$ processes, low (ideally zero) level
of radioactive contamination and ability of particle
discrimination to further reduce the background, good energy resolution,
large mass, and stability of operation over long (several years) time.

In the next Section we describe the development of large volume
cadmium tungstate crystal enriched in $^{116}$Cd. The
characterization of $^{116}$CdWO$_4$ crystal scintillators
produced from the crystal is presented in Section 3. The low
background detector with the enriched scintillators is described
in Section 4. The data of low background measurements, both in
scintillation mode and with the help of ultra-low background high
purity germanium (HPGe) $\gamma$ ray spectrometry, were analyzed
to estimate radioactive contamination of the crystal
scintillators. Finally we have simulated the detector response to
the double $\beta$ decay processes in $^{116}$Cd and estimated the
sensitivity to the neutrinoless double $\beta$ decay of
$^{116}$Cd.

\section{Development of $^{116}$CdWO$_4$ crystal scintillators}

The production of crystal scintillators from enriched materials should
satisfy some specific demands: minimal loss of expensive
isotopically enriched material, high yield of crystals, prevention
of radioactive contamination.

\subsection{Contamination of enriched $^{116}$Cd}

To produce CdWO$_4$ crystals with good scintillation
characteristics, it is necessary to control and minimize
the contamination of the initial materials for the crystal growth at a
level of $0.1-1$ ppm for a range of elements. The most dangerous
impurities which deteriorate optical and scintillation quality of
cadmium tungstate crystals are transition metals (Ti, V, Cr, Mn,
Fe, Co, Ni, Cu). Concentration of Ca, Al, Zn, Ag should also
be under control. To obtain radiopure scintillators, one should
avoid contamination by radioactive elements K, Th, U, Ra.

Samples of enriched cadmium of rather different purity grade were
used in the present work. Some parts of the enriched material
remained after the purification to produce $^{116}$CdWO$_4$
crystals for the Solotvina experiment, another part of $^{116}$Cd
oxide was extracted from the residual after the $^{116}$CdWO$_4$
crystal growth for the experiment \cite{Dan03a} (see subsection
2.2). One portion of $^{116}$Cd (316 g) was previously used in
experiment to search for double beta decay of $^{116}$Cd to the
excited states of daughter nuclei \cite{Pie94}.

The contamination of the enriched cadmium was measured with the
help of High Resolution Inductively Coupled Plasma Mass
Spectrometric analysis (Thermo Fisher Scientific ELEMENT2).
Samples' dissolution was performed by microwave treatment
according to the EPA 3052 method. Since the solids were not
completely dissolved after the microwave digestion treatment, the
supernatant of samples was analyzed by ICP MS in order to
calculate the amount of dissolved sample. To enhance the
sensitivity to Th and U, the sample solution was treated in an
extraction chromatographic system in order to separate analytes
from the matrix. This procedure allowed to reduce the dilution
factor before analysis up to a value of 150 and to minimize the
isobaric interferences between Th and WO$_3$ ions.

Potassium and iron were measured in High Resolution mode.
Concentrations were calculated based on an external calibration
method. We estimate the uncertainties of the measurements as about
15\% of the given values. Thorium and uranium were measured in
Medium Resolution mode while other elements were determined in Low
Resolution  High Sensitivity mode. A semiquantitative analysis
was performed, that is a single standard solution containing some
elements at a known concentration level (10 ppb of Li, Y, Ce, Tl)
was used for calibration and quantification. The uncertainties
when working in semiquantitative mode are about 25\% of the given
concentration value. The contribution of a blank procedure was
estimated and subtracted from the samples. The analysis results
are presented in Table \ref{tab:01}.

\begin{table}[htb]
\caption{Contamination of initial enriched $^{116}$Cd (the average
value of several samples) and $^{116}$CdWO$_4$ crystal analyzed by
ICP-MS analysis. The concentrations of impurities in the samples
of $^{116}$Cd oxide and the averaged contaminations of initial
$^{116}$Cd are normalized on the mass of cadmium.}

\begin{center}
\begin{tabular}{llll}
 \hline
  Element       & \multicolumn{3}{c}{Concentration (ppm)} \\
 \cline{2-4}
  ~             & Samples of $^{116}$Cd & Averaged              & $^{116}$CdWO$_4$\\
  ~             & and $^{116}$CdO       & contamination         & crystal \\
  ~             & ~                     & of initial $^{116}$Cd & ~ \\
  \hline
  Mg            & $<0.05-370$        & $12$                     & $<5$ \\
  Al            & $<0.5-58$          & $8$                      & $<15$ \\
  K             & $<5-17$            & $<10$                    & $5$ \\
  Ca            & $<6-37$            & 19                       & $<150$ \\
  Ti            & $<0.1-7$           & 1.5                      & $0.8$ \\
  V             & $<0.005-0.16$      & 0.01                     & $<0.15$ \\
  Cr            & $<0.05-26$         & 2                        & $<0.8$ \\
  Mn            & $<0.1-3$           & 0.7                      & $<0.3$ \\
  Fe            & $<0.07-150$        & 14                       & $3$ \\
  Co            & $<0.003-0.2$       & 0.06                     & $0.06$ \\
  Ni            & $<0.05-18$         & 1.8                      & $<0.5$ \\
  Cu            & $<0.05-10$         & 7                        & $<50$ \\
  Zn            & $<0.5-850$         & 42                       & $<0.5$ \\
  Ag            & $<0.005-1$         & 0.08                     & $<0.1$ \\
  Ba            & $<0.05-9$          & 0.3                      & $<0.2$ \\
  Th            & $<0.001-0.003$     & $<0.02$                  & $<0.00003$ \\
  U             & $<0.001-0.005$     & $<0.02$                  & $<0.0004$ \\
 \hline
\end{tabular}
\end{center}
 \label{tab:01}
\end{table}

\subsection{Recovery of enriched cadmium $^{116}$Cd from cadmium tungstate
crystalline residue}

A rest after $^{116}$CdWO$_4$ crystals growing for the Solotvina
experiment was decomposed to extract enriched cadmium. Cadmium
tungstate is not soluble in acids and alkalis. Molten sodium
carbonate (Na$_2$CO$_3$, TraceSelect, 99.9999\%) was used as a
solvent to decompose cadmium tungstate crystalline rest. A mixture
of $^{116}$CdWO$_4$ and Na$_2$CO$_3$ in the mass proportion $1:1$
was prepared, which corresponds to approximately $3.3:1$ molar
ratio. Then the mixture was heated in a platinum crucible at the
temperature $950~^{\circ}$C over 4 hours. As a result the
following reaction has occurred:

\begin{center}
 Na$_2$CO$_3$ + CdWO$_4~\to~$Na$_2$WO$_4$ + CdO.
\end{center}

Sodium tungstate (Na$_2$WO$_4$) has the melting point of
$696~^{\circ}$C. Therefore this compound was in the liquid phase
at the temperature of the reaction, while cadmium oxide is
insoluble both in the molten Na$_2$CO$_3$ and Na$_2$WO$_4$ salts
and remained in the form of sediment. After the full decomposition of the
cadmium tungstate the crucible was cooled down to room
temperature. Solid Na$_2$CO$_3$ and Na$_2$WO$_4$ were dissolved in
hot deionized water. Finally the precipitation of the enriched
cadmium oxide was rinsed and dried.

\subsection{Purification of enriched cadmium by vacuum distillation and filtering}

Most contaminated samples of enriched $^{116}$Cd were purified by
distillation through getters in the National Science Center
``Kharkiv Institute of Physics and Technology'' (Kharkiv, Ukraine)
\cite{Kov11}.

\subsection{Synthesis of $^{116}$CdWO$_4$ compound}

The powder to grow the $^{116}$CdWO$_4$ crystal was produced by
the NeoChem company (Moscow, Russia). All the operations were
carried out by using quartz or polypropylene lab-ware, materials
with low level of radioactive contaminations. Reagents of high
purity grade (concentration of any metal less than 0.01 ppm) were
used. Water, acids and ammonia were additionally distilled by
laminar evaporation in quartz installation. The high cost of the
enriched $^{116}$Cd limits the choice of the methods for its
additional purification. Recrystallization methods, typically used
for the cadmium salts purification, cannot be applied due to the low
outcome of the final product ($<85\%$). Therefore, after
dissolving the metallic cadmium in nitric acid, the purification was
realized by coprecipitation on a collector. Additional
recrystallization was performed to purify ammonium para-tungstate
used as tungsten source. Solutions of cadmium nitrate and ammonium
para-tungstate were mixed and then heated to precipitate cadmium
tungstate:

\begin{center}
Cd(NO$_3)_2$ + (NH$_4)_2$WO$_4$ = CdWO$_4$ + 2NH$_4$NO$_3$.
\end{center}

\noindent Then the $^{116}$CdWO$_4$ sediment was rinsed and
filtered. Finally the $^{116}$CdWO$_4$ compound was dried and
annealed.

\subsection{Growth of $^{116}$CdWO$_4$ crystal and production of scintillation elements}

The $^{116}$CdWO$_4$ crystal was grown by the low-thermal-gradient
Czochralski technique \cite{Pav92,Bor01,Gal09} in a platinum
crucible. Under low-thermal-gradient conditions the temperature of
the melt was close to the melting point of cadmium tungsten
crystal, which is $\approx1270~^{\circ}$C. A crystal boule with
mass of 1868 g (see Fig. 1, left) was grown from 2139 g of the
initial $^{116}$CdWO$_4$ charge (87\% of initial charge).

\nopagebreak
\begin{figure}[htbp]
\begin{center}
\resizebox{0.485\textwidth}{!}{\includegraphics{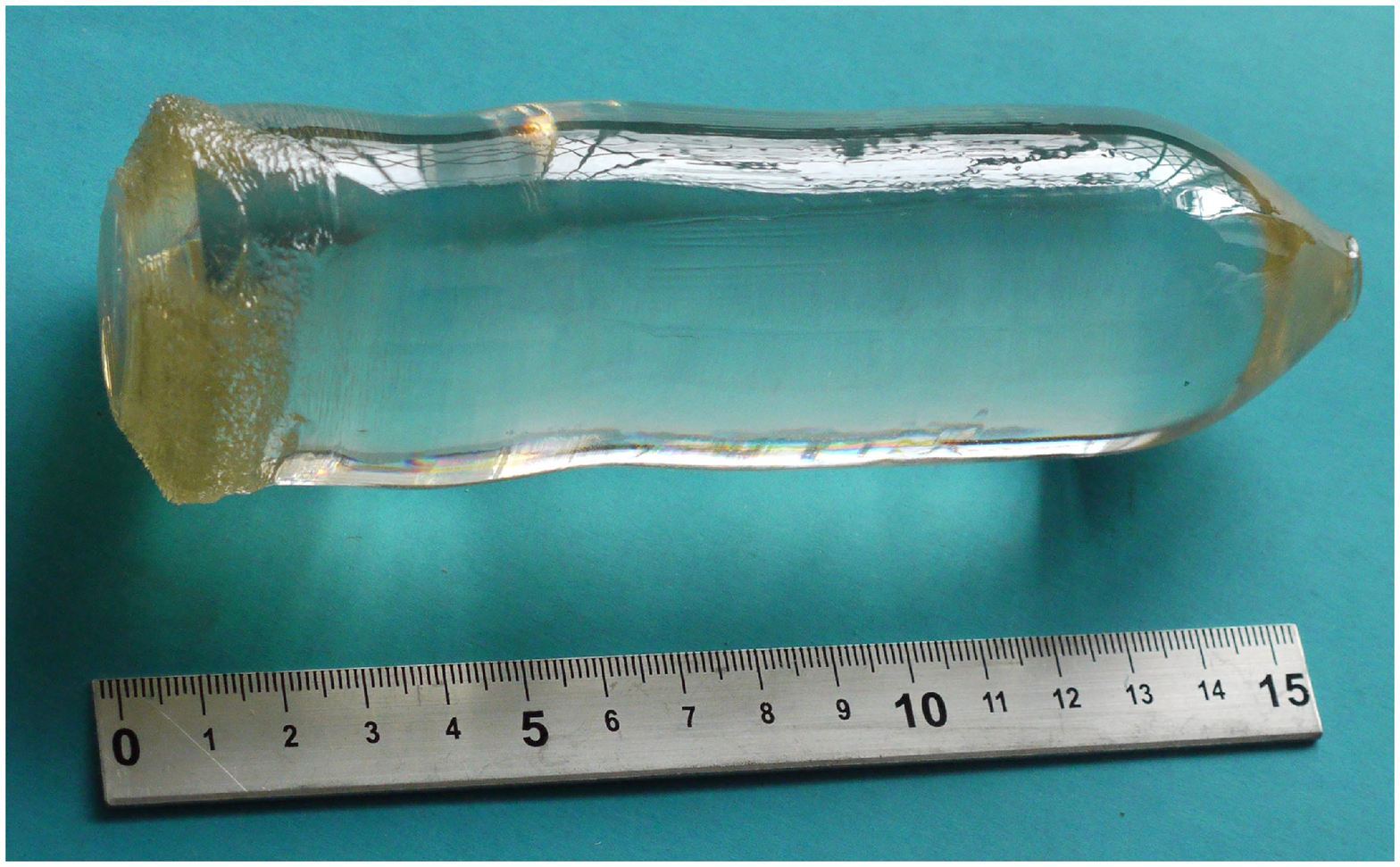}}
\resizebox{0.45\textwidth}{!}{\includegraphics{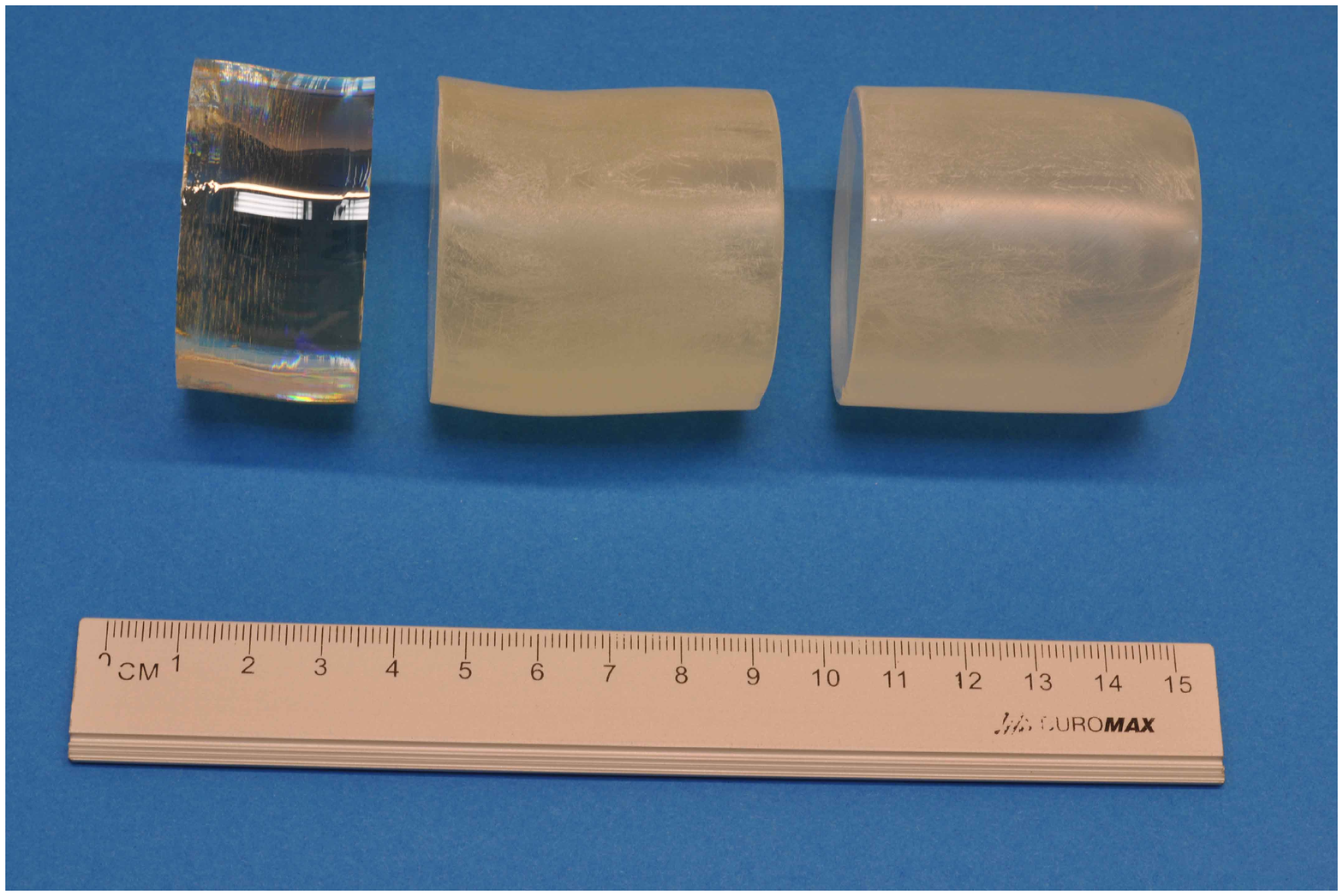}}
 \caption{(Color online) Left: Boule of enriched $^{116}$CdWO$_4$ crystal. The
conic part of the boule is the beginning of the crystal growth.
Right: Crystal samples cut from the boule:
$\approx\oslash45\times46.7$~mm, 586.2 g, No. 1 (right);
$\approx\oslash45\times46.1$~mm, 589.3 g, No. 2 (middle);
$\approx\oslash45.7\times25.1$~mm, 325.6 g, No. 3 (left).}
\end{center}
 \label{fig:01}
\end{figure}

Three near cylindrical shape crystal scintillators
($\oslash45\times46.7$~mm, 586.2 g, No. 1;
$\oslash45\times46.1$~mm, 589.3 g, No. 2;
$\oslash45.7\times25.1$~mm, 325.6 g, No. 3) have been cut from the
crystal boule (see Fig. 1, right). The side surface of the
crystals No.1 and 2 was diffused with the help of grinding paper
to reach uniformity of scintillation light collection, which is
important to improve energy resolution of the detector.

\section{Characterisation of $^{116}$CdWO$_4$ crystal scintillators}

\subsection{Isotopic composition of cadmium in the $^{116}$CdWO$_4$ crystal}

The isotopic composition of the cadmium in the enriched
$^{116}$CdWO$_4$ crystal was measured with the help of the High
Resolution Inductively Coupled Plasma Mass Spectrometric analysis.
Results of the analysis are presented in Table \ref{tab:02}. The
absolute isotope abundance for $^{116}$Cd is $82.2\%$, while
$\beta$ active $^{113}$Cd has an absolute isotope abundance of
$2.14\%$. The atomic weight of cadmium in the $^{116}$CdWO$_4$
crystal is 115.3 as compared to the table value of
$112.411\pm0.008$ \cite{Wie11}. The atomic weight of
$^{116}$CdWO$_4$ molecule is 363.1.

\begin{table}[htb]
\caption{The absolute isotopic composition of cadmium in the
$^{116}$CdWO$_4$ crystal (\%).}
\begin{center}
\begin{tabular}{lll}
 \hline
  Atomic number & Enriched $^{116}$Cd   & Natural cadmium \cite{Ber11}\\
  \hline
  106           & $0.11\pm0.01$         & $1.25\pm0.06$ \\
  108           & $0.10\pm0.01$         & $0.89\pm0.03$ \\
  110           & $1.80\pm0.05$         & $12.49\pm0.18$ \\
  111           & $2.00\pm0.05$         & $12.80\pm0.12$ \\
  112           & $4.35\pm0.04$         & $24.13\pm0.21$ \\
  113           & $2.14\pm0.06$         & $12.22\pm0.12$ \\
  114           & $7.30\pm0.06$         & $28.73\pm0.42$ \\
  116           & $82.2\pm0.1$          & $7.49\pm0.18$ \\
   \hline
\end{tabular}
\end{center}
 \label{tab:02}
\end{table}

\subsection{Light transmission}

\begin{figure*}[htb]
\begin{center}
\resizebox{0.7\textwidth}{!}{\includegraphics{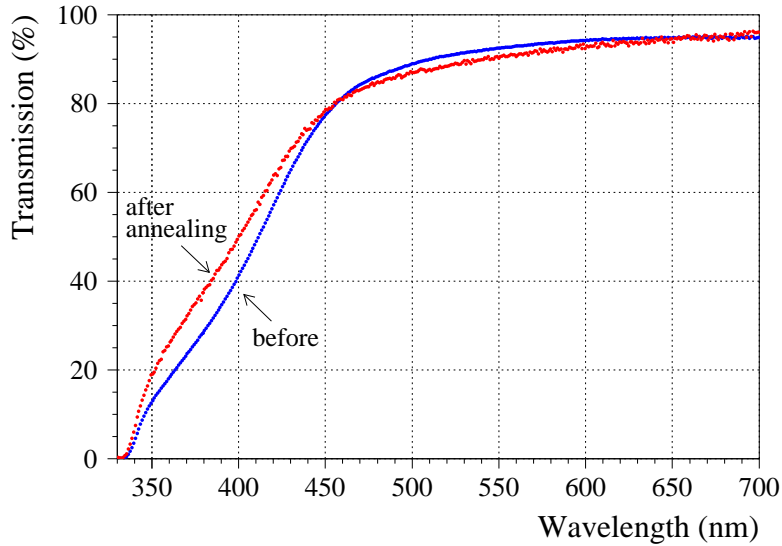}}
\end{center}
\caption{(Color online) The optical transmission curve of
$^{116}$CdWO$_4$ crystal No. 2 before and after annealing measured
with 2.6~mm sample in reference beam.}
 \label{fig:02}
\end{figure*}

The transmittance of the $^{116}$CdWO$_4$ crystal scintillators
was measured in the spectral range $330-700$ nm using a PERKIN
ELMER UV/VIS spectrometer Lambda 18. A thin (2.6 mm) sample of the
$^{116}$CdWO$_4$ crystal was placed in the reference beam of the
instrument to correct the reflection losses. The results of the
optical transmission measurements for the $^{116}$CdWO$_4$ crystal
No. 2 shown in Fig. \ref{fig:02} demonstrate that the material
exhibits reasonable transmission properties in the relevant
wavelength range $400-600$ nm of the CdWO$_4$ emission spectrum.

From the data of the transmission measurements we have derived the
attenuation length of the material $31\pm5$ cm at the wavelength
of CdWO$_4$ emission maximum 480 nm \cite{Bar06}. At 400 nm the
attenuation lengths of the samples are 5.8 cm (No. 1), 5.2 cm (No.
2) and 4.1 cm (No. 3). One can explain the decrease of
transmittance for the samples distant from the beginning of the
crystal growth by the well known effect of defects increase during
CdWO$_4$ crystal growth.

After the low background measurements (see below subsection 4.2)
the crystals No. 1 and 2 were annealed at the temperature
$\approx870~^{\circ}$C over 55 hours. The annealing improved
transmittance of the samples in the region of wavelengths
$350-420$ nm on $10-40~\%$ (see Fig. \ref{fig:02}).

\subsection{Energy resolution}

\begin{figure*}[htb]
\begin{center}
\resizebox{0.6\textwidth}{!}{\includegraphics{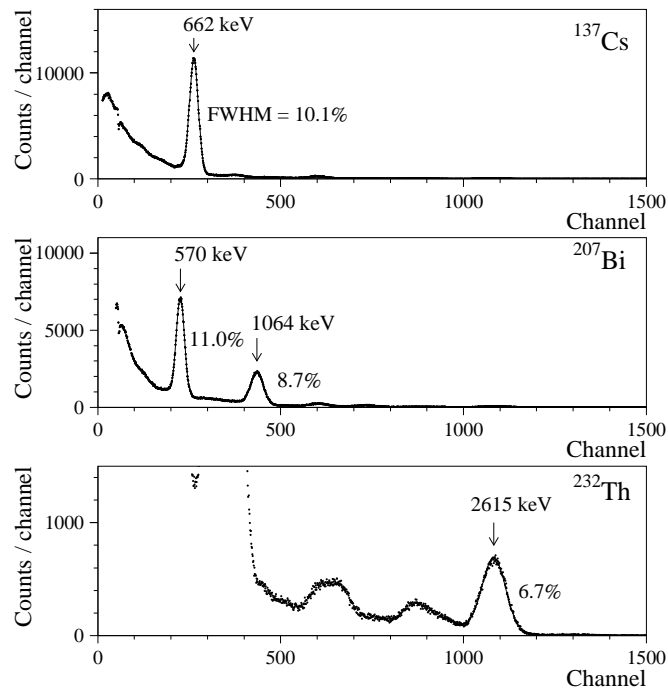}}
\end{center}
\caption{The energy spectra of $^{137}$Cs,
$^{207}$Bi and $^{232}$Th $\gamma$ quanta measured by the
$^{116}$CdWO$_4$ scintillation crystal No. 2.}
 \label{fig:03}
\end{figure*}

To measure the scintillation properties, the samples No. 1 and 2
were optically coupled with the help of Dow Corning Q2-3067
optical couplant to 3" photomultiplier (PMT) Philips XP2412. To
improve scintillation light collection, the crystals were wrapped
by a few layers of polytetrafluoroethylene (PTFE) tape. The
measurements were carried out with 10 $\mu$s shaping time of ORTEC
575 spectroscopy amplifier to collect most of the charge from the
anode of the PMT. The detectors were irradiated by $\gamma$ quanta
of $^{137}$Cs, $^{207}$Bi and $^{232}$Th sources. Fig.
\ref{fig:03} shows the pulse amplitude spectra measured with the
$^{116}$CdWO$_4$ crystal scintillator No. 2. The energy resolution
11.1\% and 10.1\% (FWHM) were obtained for the 662 keV $\gamma$
line of $^{137}$Cs with the detectors No. 1 and 2, respectively.
The energy resolution for 2615 keV $\gamma$ line of $^{232}$Th
source is 7.1\% and 6.7\% for the scintillators No.~1 and 2,
respectively.

After the annealing at high temperature the energy resolution of
the crystals was improved (see subsection 4.2).

\section{Low background measurements, results and discussion}

\subsection{Measurements with ultra-low background HPGe $\gamma$ ray spectrometry}

\begin{figure*}[htb]
\begin{center}
\resizebox{0.8\textwidth}{!}{\includegraphics{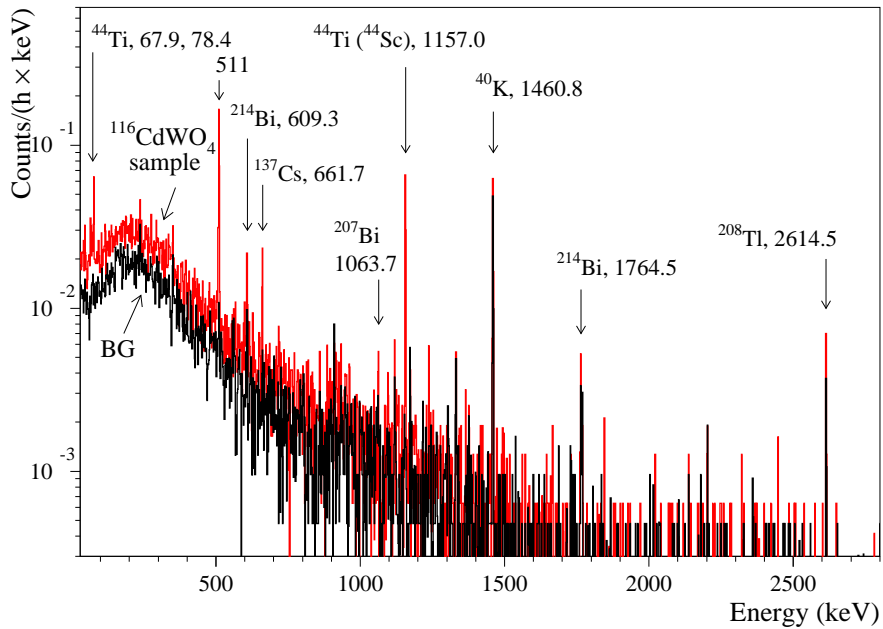}}
\end{center}
\caption{(Color online) Energy spectra measured with the 325.6 g
$^{116}$CdWO$_4$ sample over 788 h and without sample over 1046 h
(BG) by ultra-low background HPGe $\gamma$ spectrometer. Some
excess of the spectrum accumulated with the $^{116}$CdWO$_4$
sample is due to an accidental contamination of the HPGe detector
by radioactive $^{44}$Ti. The energy of the $\gamma$ lines are in
keV.} \label{fig:04}
\end{figure*}

The $^{116}$CdWO$_4$ crystal No.~3 was measured for 788 h with the
ultra-low background HPGe $\gamma$ ray spectrometer GeCris (volume
468 cm$^3$, 120\% relative efficiency). The background data were
accumulated over 1046 h (see Fig. \ref{fig:04}). In order to
determine the radioactive contamination of the sample, the
detection efficiencies were calculated using Monte Carlo
simulation based on the GEANT4 software package  \cite{GEANT4}.
Peaks in the spectra are due to the naturally occurring
radionuclides of the uranium and thorium chains and $^{40}$K. Only
upper limits could be obtained for corresponding activities. We
have detected low contamination by $^{137}$Cs and $^{207}$Bi in
the crystal on the level of 2.1(5) mBq/kg and 0.6(2) mBq/kg,
respectively.  In addition, we have observed peaks of $^{44}$Ti
(67.9 keV and 78.4 keV, the half-life of $^{44}$Ti is 60 yr) and
its daughter $^{44}$Sc (1157.0 keV) in the data accumulated with
the $^{116}$CdWO$_4$ crystal. However, the peaks are due to
contamination of the HPGe detector (not the crystal scintillator
sample) by $^{44}$Ti. Indeed the detector, before the run with the
$^{116}$CdWO$_4$ sample, has been used to measure a sample of
titanium with rather high activity of $^{44}$Ti. Limits on mBq/kg
level were obtained for other potential contaminations. The
results of the measurements are presented in Table \ref{tab:03}.

\subsection{Low background detector system with $^{116}$CdWO$_4$ crystal scintillators}

\begin{figure*}[htb]
\begin{center}
\resizebox{0.7\textwidth}{!}{\includegraphics{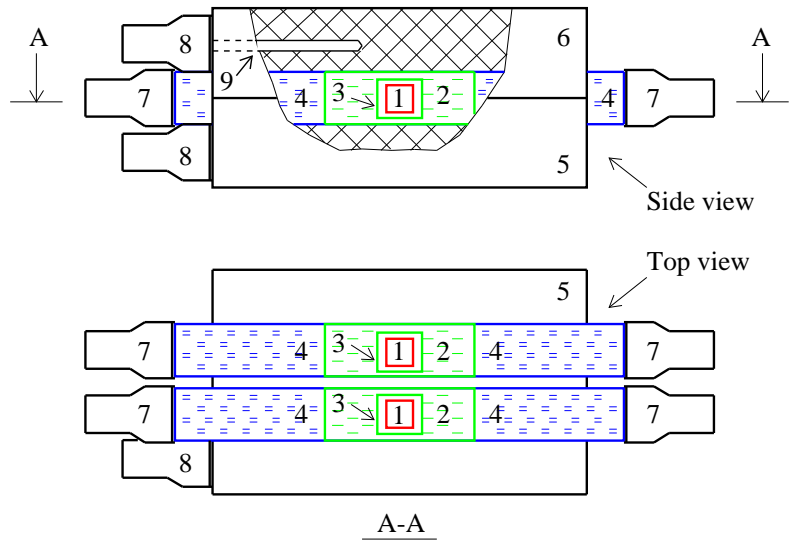}}
\end{center}
\caption{(Color online) Low background scintillation detector.
 (1) $^{116}$CdWO$_4$ crystals scintillators, (2) plastic scintillator light-guides,
(3) cavities in the light-guides filled by liquid scintillator,
(4) quartz light-guides, (5, 6) plastic scintillators, (7) four
PMTs Hamamatsu R6233MOD, (8) two PMTs ETL 9302FLA, (9) channel
$\oslash14\times200$ mm in the plastic scintillator 6 to insert
radioactive sources.}
 \label{fig:05}
\end{figure*}

The $^{116}$CdWO$_4$ crystal scintillators No.~1 and 2 were fixed
inside the cavities $\oslash47\times61$ mm in the central part of
the polystyrene based plastic scintillator light-guides (UPS923A,
Amcrys-H, Ukraine), 70 mm in diameter and 194 mm in length (a
schematic view of the set-up is presented in Fig. \ref{fig:05}).
The cavities were filled with liquid scintillator (LS-221,
Institute for Scintillation Materials, Kharkiv, Ukraine) which
does not affect the polystyrene scintillator. The scintillating
light-guides act as active veto. A significant difference of
CdWO$_4$ pulse-shape (effective average decay time is 13 $\mu s$
\cite{Bar06}) in comparison to much faster plastic and liquid
scintillators signals (few nanoseconds) offers the possibility to
exploit the discrimination of the light-guide signals. Each
plastic light-guide was optically connected on opposite sides to
two high purity quartz light-guides $\oslash70\times200$ mm each.
Two low radioactive 3" diameter PMTs Hamamatsu R6233MOD viewed
each detector from opposite sides. The light-guides are wrapped by
a few layers of PTFE reflection tape. All the optical contacts
between the light-guides and PMTs were provided by Dow Corning
Q2-3067 optical couplant. The detectors with the $^{116}$CdWO$_4$
crystals were placed between two polystyrene based plastic
scintillators (UPS89, Amcrys-H, Ukraine) $500\times300\times120$
mm. A channel $\oslash14\times200$ mm was made (in the middle of
the upper plastic scintillator of its width, 51 mm from the above
edge of the plastic) to insert radioactive sources. Two low
background 3" diameter PMTs ETL 9302FLA were optically connected
to the plastic scintillators.

The detector system was installed deep underground in the low
background DAMA/R\&D set-up at the Gran Sasso National
Laboratories of the INFN (Italy). The detector system was
surrounded by Cu bricks and sealed in a low radioactive air-tight
Cu box continuously flushed by high purity nitrogen gas (stored
deeply underground for a long time) to avoid the presence of
residual environmental radon. The Cu box was surrounded by a
passive shield made of high purity Cu, 10 cm of thickness, 15 cm
of low radioactive lead, 1.5 mm of cadmium and 4 to 10 cm of
polyethylene/paraffin to reduce the external background. The
shield was contained inside a Plexiglas box, also continuously
flushed by high purity nitrogen gas.

\begin{figure*}[htb]
\begin{center}
\resizebox{0.77\textwidth}{!}{\includegraphics{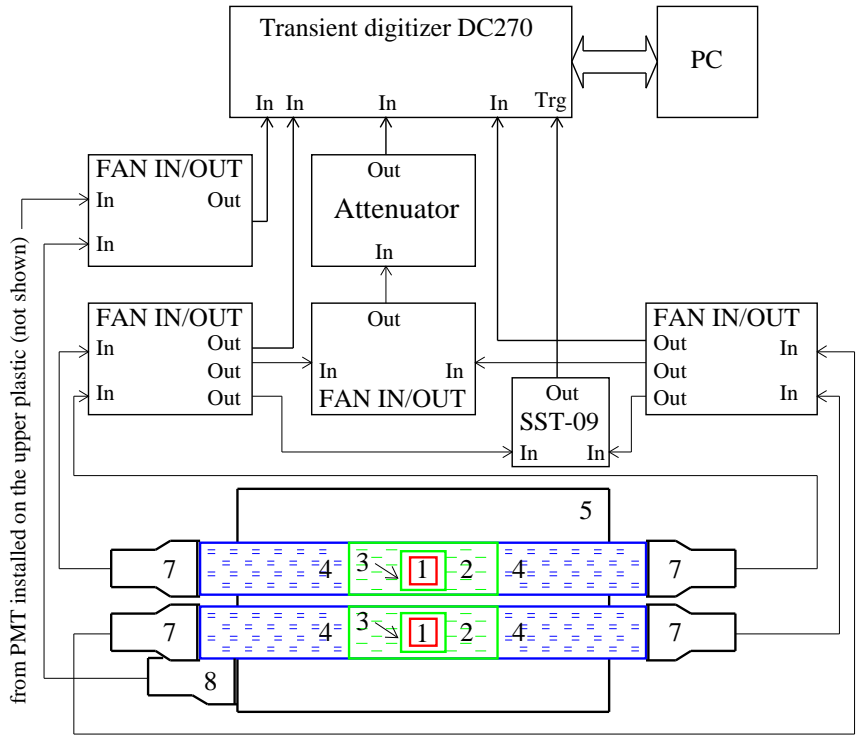}}
\end{center}
\caption{(Color online) Schema of the electronic chain (details of
the low background detector are denoted in Fig. 5). (FAN IN/OUT)
Linear FAN-IN/FAN-OUT, LeCroy Model 428F; (SST-09) home made
electronic unit to provide triggers for cadmium tugstate
scintillation signals; (Attenuator) Dual attenuator, CAEN model No
110; (PC) personal computer. Inputs and outputs of the electronic
units are denoted as "In" and "Out", respectively; "Trg" denotes a
trigger input of the transient digitizer.}
 \label{fig:06}
\end{figure*}

An event-by-event data acquisition system based on a 1 GS/s 8 bit
transient digitizer (Acqiris DC270) records the time of each event
and the pulse shape over a time window of 100 $\mu$s from the
$^{116}$CdWO$_4$ detectors (the sum of the signals from two PMTs),
the plastic scintillator, and the sum of signals from the
$^{116}$CdWO$_4$ scintillators attenuated to provide an energy
scale up to $\approx10$ MeV (the electronic chain of the detector
system is briefly summarized in Fig. \ref{fig:06}). Taking into
account the slow kinetics of the CdWO$_4$ scintillation decay, the
sampling frequency of the transient digitizer was set to 20 MS/s.
An especially developed electronic unit (SST-09) provides triggers
for cadmium tugstate scintillation  signals. The unit rejects PMT
noise, plastic scintillator light-guide signals and CdWO$_4$
events with large admixture of the plastic. Further rejection of
the plastic overlaps can be realized off line by the pulse-shape
analysis described below.

The energy scale and resolution of the detector system was tested
with $^{22}$Na, $^{60}$Co, $^{137}$Cs and $^{228}$Th $\gamma$
sources. The energy resolution of the $^{116}$CdWO$_4$ detectors
before the annealing can be described by functions:
FWHM$_{\gamma1}=\sqrt{10.5\times E_{\gamma}}$, and
FWHM$_{\gamma2}=\sqrt{540+8.5\times E_{\gamma}}$  for the
detectors No. 1 and 2, respectively (here $E_{\gamma}$ and
FWHM$_{\gamma}$ are in keV). The energy scale and resolution of
the detectors were tested once during the measurements and at the
end of the experiment with the help of $^{22}$Na $\gamma$ source
inserted into the set-up through the special channel without
switching off the high voltage of the PMTs. We have observed
neither shift of the energy scale nor degradation of the energy
resolution of the detectors during almost 2.5 months of low
background measurements.

\begin{figure*}[htb]
\begin{center}
\resizebox{0.7\textwidth}{!}{\includegraphics{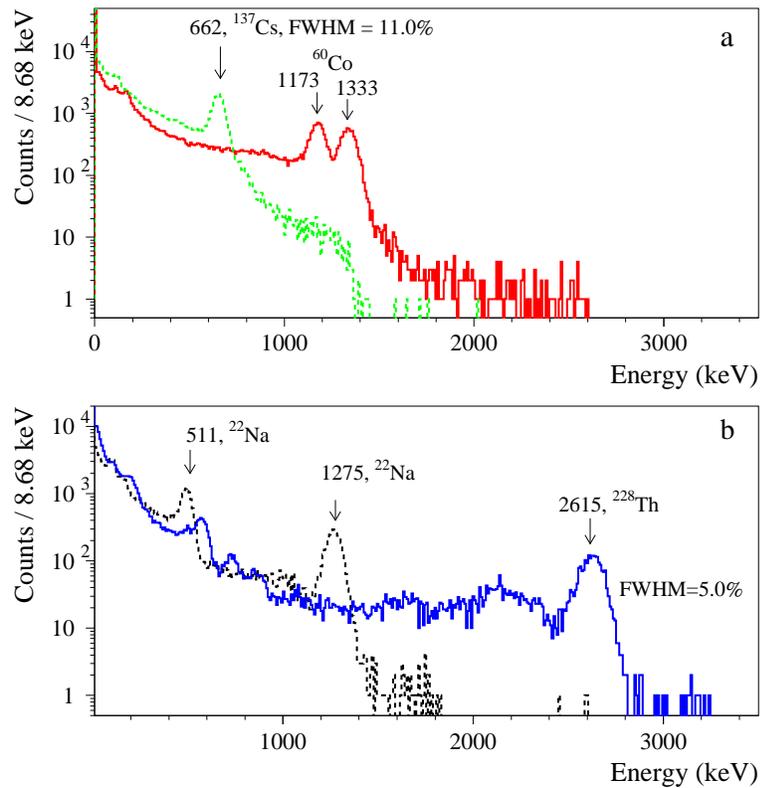}}
\end{center}
\caption{(Color online) Energy spectra accumulated by the
$^{116}$CdWO$_4$ detector No. 2 with $^{137}$Cs, $^{60}$Co (a),
$^{22}$Na and $^{228}$Th $\gamma$ sources (b) in the low
background set-up after annealing of the crystals. Energies of
$\gamma$ lines are in keV.}
 \label{fig:07}
\end{figure*}

As a result of the annealing of the $^{116}$CdWO$_4$ crystal
scintillators performed after the low background measurements, the
energy resolution of the detectors for the 2615 keV $\gamma$ line
of $^{228}$Th was improved from 6.9\% to 5.3\% for the detector
with the crystal No. 1, and from 6.2\% to 5.0\% for the detector
No. 2. Energy spectra accumulated with the $^{116}$CdWO$_4$
detector No. 2 after the annealing are presented in Fig.
\ref{fig:07}.

\subsection{Data analysis}

\begin{figure*}[htb]
\begin{center}
\resizebox{0.8\textwidth}{!}{\includegraphics{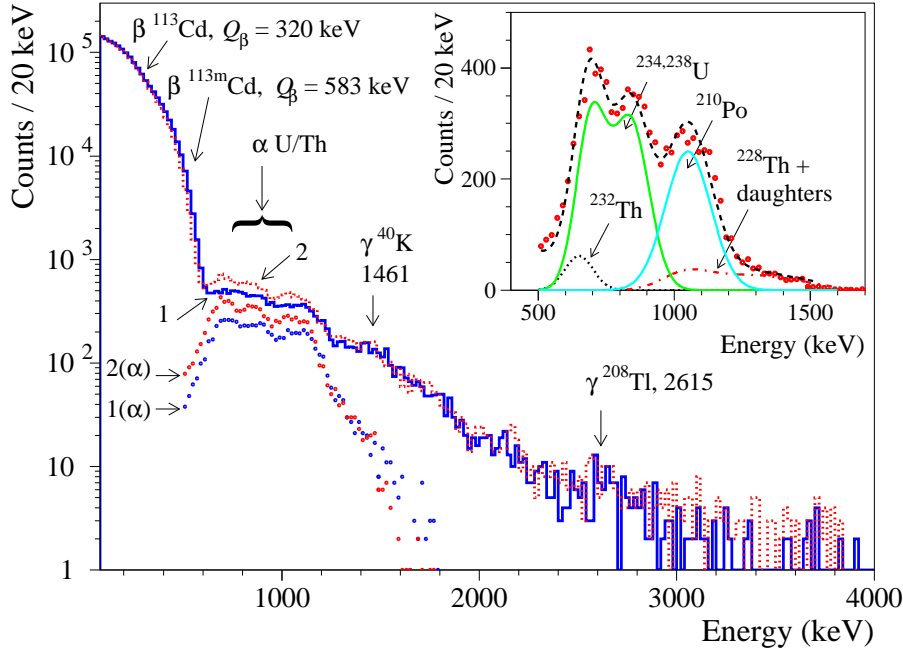}}
\end{center}
\caption{(Color online) The energy spectra accumulated with the
$^{116}$CdWO$_4$ crystal sintillators No. 1 and 2 in the low
background DAMA/R\&D set-up over 1727 h. The energy spectra of
$\alpha$ events selected by the pulse-shape discrimination (see
text) are also shown. Some difference in the data in the energy
region $0.6-1.2$ MeV visible both in the raw data and in the
spectra of $\alpha$ events is due to higher $\alpha$ activity of
U/Th daughters' traces in the crystal No. 2. In the inset, the
$\alpha$ spectrum of the detector No. 2 is depicted together with
the model, which includes $\alpha$ decays from $^{238}$U and
$^{232}$Th families.} \label{fig:08}
\end{figure*}

The energy spectra accumulated with the $^{116}$CdWO$_4$ detectors
over 1727 h of low background measurements are presented in Fig.
\ref{fig:08}. The spectra are rather similar with a small
difference in the region $0.6-1.2$ MeV where $\alpha$ particles
from U/Th are observed (see Subsection 4.3.1 describing a
pulse-shape discrimination to select $\alpha$ particles). The
counting rate of 0.26 count/s in the energy interval $0.08-0.6$
MeV is mainly due to the decay of $^{113}$Cd ($Q_{\beta}=320$ keV,
$T_{1/2}=8.04\times10^{15}$ yr) with the activity ($0.10\pm0.01$)
Bq/kg\footnote{The activity is calculated on the basis of the
isotopic abundance and the half-life of $^{113}$Cd.} and
$^{113m}$Cd ($Q_{\beta}=583$ keV, $T_{1/2}=14.1$ yr) with the
activity ($0.46\pm0.02$) Bq/kg.

Contributions to the background above the energy 0.6 MeV were
analyzed by the pulse-shape discrimination and the time-amplitude
techniques.

\subsubsection{Pulse-shape discrimination}

\begin{figure*}[htb]
\begin{center}
\resizebox{0.7\textwidth}{!}{\includegraphics{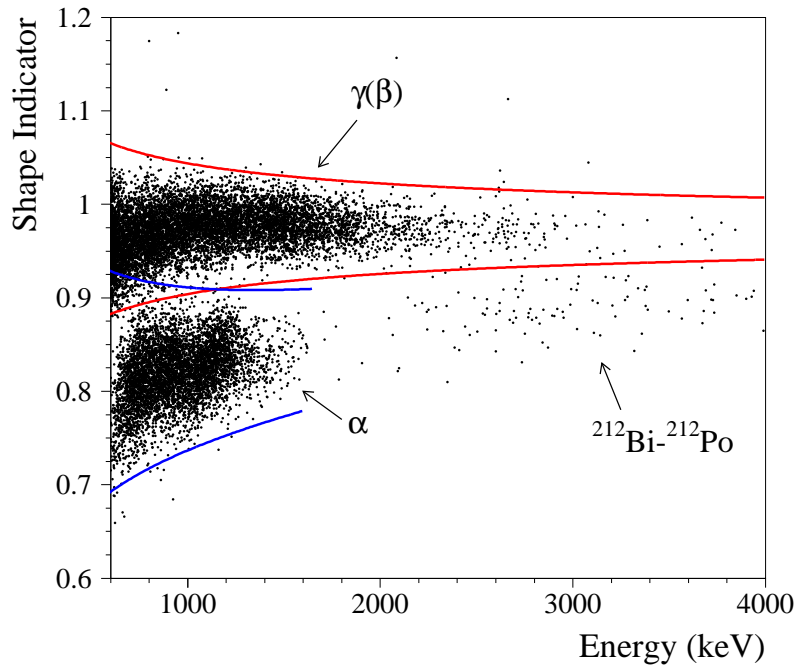}}
\end{center}
\caption{(Color online) Shape indicators (see text) versus energy
for background exposition over 1727 h with $^{116}$CdWO$_4$
crystal scintllator No. 1 in the low background set-up. Three sigma
intervals for shape indicator values corresponding to $\gamma$
quanta ($\beta$ particles) and $\alpha$ particles are drawn.}
 \label{fig:09}
\end{figure*}

To select $\gamma$ quanta ($\beta$ particles) and $\alpha$
particles, the data of the low background measurements were
analyzed by using the optimal filter method proposed by E.~Gatti
and F.~De Martini \cite{Gat62} (see also \cite{Faz98} where the
analysis was developed for CdWO$_4$ crystal scintillators). For
each experimental signal, the numerical characteristic of its
shape (shape indicator, $SI$) was defined as $SI=\sum f(t_k)\times
P(t_k)/\sum f(t_k)$, where the sum is over time channels $k$,
starting from the origin of the signal and up to 30 $\mu$s,
$f(t_k)$ is the digitized amplitude (at the time $t_k$) of a given
signal. The weight function $P(t)$ is defined as:
$P(t)=\{\overline{f}_\alpha (t)-\overline{f}_\gamma
(t)\}/\{\overline{f} _\alpha (t)+\overline{f}_\gamma (t)\}$, where
the reference pulse shapes $\overline{f}_\alpha (t)$ and
$\overline{f}_\gamma (t)$ are taken from \cite{Bar06}. The
pulse-shape discrimination of the events accumulated in the low
background measurements with the $^{116}$CdWO$_4$ detector No. 1
is demonstrated in Fig. \ref{fig:09}. Alpha events were selected
from the accumulated data (see the energy spectra of $\alpha$
particles in Inset of Fig. \ref{fig:08}), which allow to estimate
the total internal $\alpha$ activity of U/Th as 1.9(2) mBq/kg and
2.7(3) mBq/kg in the $^{116}$CdWO$_4$ crystal scintillators No. 1
and 2, respectively. Slightly higher $\alpha$ activity in the
sample No. 2 can be explained by accumulation of U/Th daughters'
traces in the melt during the crystal growth.

To estimate the activity of the $\alpha$ active nuclides from the
U/Th families in the crystals, the energy spectra of the $\alpha$
events were fitted in the energy interval $0.5-1.5$ MeV by using a
simple model: ten Gaussian functions to describe $\alpha$ peaks of
$^{232}$Th (and its daughters: $^{228}$Th, $^{224}$Ra, $^{220}$Rn,
$^{216}$Po, $^{212}$Bi), $^{238}$U (and its daughters: $^{234}$U,
$^{230}$Th, $^{210}$Po) plus exponential function (to describe
background). Fit of the $\alpha$ spectrum accumulated with the
detector No. 2 is shown in Inset of Fig. \ref{fig:08}. Because of
the worse energy resolution for $\alpha$ particles in comparison
to $\gamma$ quanta in CdWO$_4$ scintillation detectors
\cite{Dan03b}, we conservatively give limits on activities of
$^{232}$Th and $^{238}$U, $^{230}$Th, $^{210}$Po (expected to be
not in equilibrium with $^{238}$U) in the $^{116}$CdWO$_4$
scintillators. The data obtained from the fits are presented in
Table \ref{tab:03}.

In addition, the optimal filter method allows to distinguish and reject
from the data also overlapping of plastic scintillator's pulses with
$^{116}$CdWO$_4$ signals, random coincidence of events, some
events from the fast chain of decays $^{212}$Bi$\to ^{212}$Po$\to
^{208}$Pb ($^{232}$Th family).

\subsubsection{Time-amplitude analysis of $^{228}$Th and $^{227}$Ac contaminations}

Activities of $^{228}$Th ($^{232}$Th family) and $^{227}$Ac
($^{235}$U) in the $^{116}$CdWO$_4$ crystal scintillators were
estimated by the time-amplitude analysis of the events accumulated in the
low background measurements. The technique of the time-amplitude
analysis is described in detail in \cite{Dan03a,Dane95,Dane01}.

\begin{figure*}[htb]
\begin{center}
\resizebox{0.7\textwidth}{!}{\includegraphics{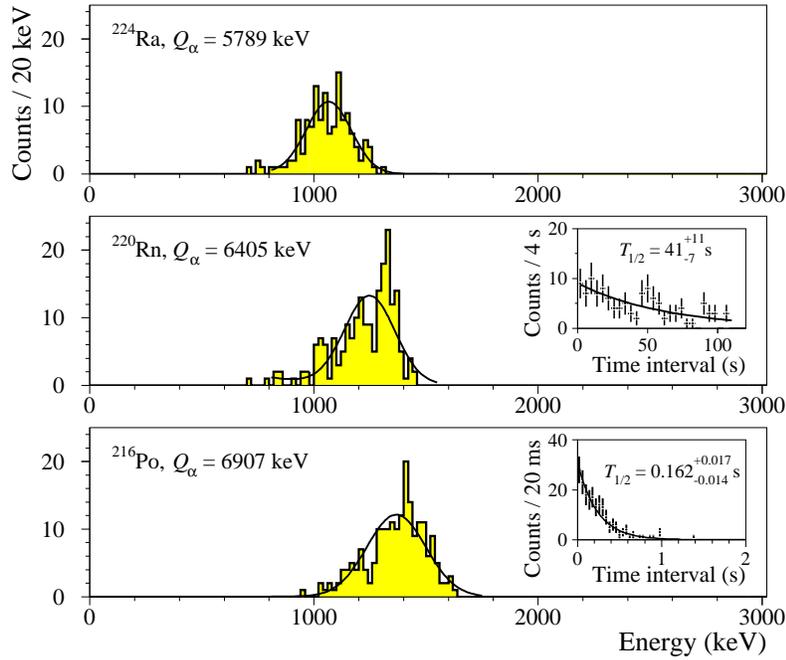}}
\end{center}
\caption{(Color online) Alpha peaks of $^{224}$Ra, $^{220}$Rn and
$^{216}$Po selected by the time-amplitude analysis from the data
accumulated during 1727 h with the $^{116}$CdWO$_4$ detector No.
1. The obtained half-lives of $^{220}$Rn ($41^{+11}_{-7}$ s) and
$^{216}$Po ($0.162^{+0.017}_{-0.014}$ s) are in agreement with the
table values (55.6 s and 0.145 s, respectively \cite{ToI98}).}
 \label{fig:10}
\end{figure*}

To determine the activity of $^{228}$Th, the following sequence of
$\alpha $ decays was selected: $^{224}$Ra ($Q_\alpha $ = $5.789$
MeV, $T_{1/2}$ = $3.66$ d) $ \to $ $^{220}$Rn ($Q_\alpha $ =
$6.405$ MeV, $T_{1/2}$ = $55.6$ s) $ \to $ $^{216}$Po ($Q_\alpha $
= $6.907$ MeV, $T_{1/2}$ = $0.145$ s) $\to $ $^{212}$Pb. The
obtained $\alpha $ peaks and the distributions of the time
intervals between events (see Fig. \ref{fig:10}) are in agreement
with those expected for $\alpha $ particles of the chain. Taking
into account the efficiencies in the time windows to select
$^{220}$Rn $\to $ $^{216}$Po $\to $ $^{212}$Pb events (94.5\%),
the activities of $^{228}$Th in the crystals No. 1 and 2 were
calculated as 0.057(7) mBq/kg and 0.062(6) mBq/kg, respectively.

By using positions of the three $\alpha$ peaks in the $\gamma$
scale of the detector we have obtained the following dependence of
$\alpha/\beta$ ratio on energy of $\alpha$ particles $E_{\alpha}$:
$\alpha/\beta=0.113(6)+0.132(10)\times10^{-4}E_{\alpha}$ in the
energy interval $5.8-6.9$ MeV ($E_{\alpha}$ is in keV).

The same approach was used to search for the chain $^{223}$Ra
($Q_\alpha = 5.979$ MeV, $T_{1/2} = 11.44$~d) $\to$ $^{219}$Rn
($Q_\alpha = 6.946$ MeV, $T_{1/2} = 3.96$~s) $\to$ $^{215}$Po
$(Q_\alpha = 7.526$ MeV, $T_{1/2} = 1.78$ ms) $\to$ $^{211}$Pb
from the $^{235}$U family. Limit on activities of $^{227}$Ac in
the crystals No. 1 and 2 was obtained on the level of $\leq 0.002$
mBq/kg.

\subsubsection{Selection of Bi--Po events}

Double pulses from the decays of  $^{214}$Bi ($Q_\beta =3.272$
MeV) $\to $ $^{214}$Po ($Q_\alpha =7.833$ MeV, $ T_{1/2}=164$ $\mu
$s) $\to $ $^{210}$Pb (daughters of $^{226}$Ra from the $^{238}$U
family) were searched for. One event in the data accumulated with
both detectors was found. Therefore, taking into account the
efficiency of the Bi-Po events selection in the time interval of
the analysis $1-75~\mu$s (26.7\%), we set limit on $^{226}$Ra
activity in the crystals No. 1 and 2 as $\leq0.005$ mBq/kg.

A summary of the radioactive contamination of the $^{116}$CdWO$_4$
crystal scintillators is presented in Table \ref{tab:03}.

\begin{table}[htbp]
\caption{Radioactive contamination of $^{116}$CdWO$_4$ crystal
scintillators. Data for CdWO$_4$ \cite{Bel07}, $^{116}$CdWO$_4$
\cite{Dan03b,Dan03a}, and $^{106}$CdWO$_4$ \cite{Bel10bc} are
given for comparison.}

\begin{center}
\begin{tabular}{llllllll}
 \hline
 Chain      & Nuclide                           & \multicolumn{6}{c}{Activity (mBq/kg)} \\
 ~          & (Sub-chain)                       & \multicolumn{3}{c}{$^{116}$CdWO$_4$}      & CdWO$_4$      & $^{116}$CdWO$_4$  & $^{106}$CdWO$_4$  \\
  ~         & ~                                 & No. 1         & No. 2         & No. 3     & \cite{Bel07}  & \cite{Dan03b,Dan03a}& \cite{Bel10bc}\\
 \hline
 ~          & $^{40}$K                          & ~             & ~             & $\leq26$  & $\leq5$       & 0.3(1)            & $\leq11$ \\
 ~          & $^{60}$Co                         & ~             & ~             & $\leq0.47$& $\leq0.4$     & ~                 & ~ \\
 ~          & $^{110m}$Ag                       & 0.06(4)       & 0.06(4)       & ~         & ~             & ~                 & ~ \\
 ~          & $^{113}$Cd                        & 100(10)       & 100(10)       & ~         &  558(4)       & $91(5)$           & 174 \\
 ~          & $^{113m}$Cd                       & 460(20)       & 460(20)       & ~         & $\leq3.4$     & $1.1(1)$          & $112 000(5 000)$ \\
 ~          & $^{137}$Cs                        & ~             & ~             & 2.1(5)    & $\leq0.3$     & $0.43(6)$         & ~ \\
 ~          & $^{207}$Bi                        & ~             & ~             & 0.6(2)    &               & ~                 & 1.3(3) \\
 ~          & ~                                 & ~             & ~             & ~         & ~             & ~                 & ~ \\
 $^{232}$Th & $^{232}$Th                        & $\leq0.08$   & $\leq0.08$     & ~         & $\leq0.026$   & 0.053(9)          & $\leq0.1$ \\
 ~          & $^{228}$Th                        & 0.057(7)      & 0.062(6)      & $\leq2.0$ &  0.008(4)     & 0.039(2)          & 0.053(5) \\
 ~          & ~                                 & ~             & ~             &           & ~             & ~                 & ~ \\
 $^{235}$U  & $^{235}$U                         & ~             & ~             & $\leq4.0$ & ~             & ~                 & ~ \\
 ~          & $^{227}$Ac                        & $\leq0.002$   & $\leq0.002$   & ~         & 0.014(9)      & 0.0014(9)         & ~ \\
 ~          & ~                                 & ~             & ~             & ~         & ~             & ~                 & ~ \\
 $^{238}$U  & $^{238}$U                         & $\leq0.4$     & $\leq0.6$     & ~         & $\leq0.045$   & $\leq0.6$         & $\leq0.3$ \\
 ~          & $^{234m}$Pa                       & ~             & ~             & $\leq58$  & ~             & $\leq0.2$         &  \\
 ~          & $^{230}$Th                        & $\leq0.06$    & $\leq0.05$    & ~         & ~             & $\leq0.5$         & $\leq0.8$ \\
 ~          & $^{226}$Ra                        & $\leq0.005$   & $\leq0.005$   & $\leq2.6$ & $\leq0.018$   & $\leq0.004$       & $\leq0.3$ \\
 ~          & $^{210}$Pb                        & ~             & ~             & $\leq15000$ & ~             & $\leq0.4$         & ~ \\
 ~          & $^{210}$Po                        & $\leq0.4$     & $\leq0.6$     & ~         & $\leq0.063$   & ~                 & ~ \\
 ~          & ~                                 & ~             & ~             & ~         & ~             & ~                 & $\leq0.3$ \\
 \multicolumn{2}{c}{Total $\alpha$ activity}    & $1.9(2)$      & 2.7(3)        & ~         & $0.26(4)$     & 1.40(10)          & 2.1(1) \\
 \hline

\end{tabular}
\end{center}
 \label{tab:03}
\end{table}

\begin{figure*}[htb]
\begin{center}
\resizebox{0.55\textwidth}{!}{\includegraphics{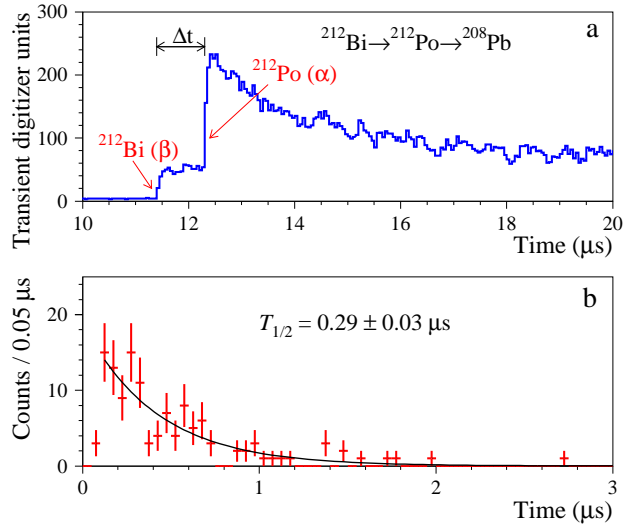}}
\end{center}
\caption{(Color online) (a) Example of $^{212}$Bi$\to
^{212}$Po$\to ^{208}$Pb event in the $^{116}$CdWO$_4$
scintillator. (b) The time distribution for the fast sequence of
$\beta $ ($^{212}$Bi) and $\alpha $ ($^{212}$Po) decays selected
by the pulse-shape and the front edge analyses from the background
data accumulated with the $^{116}$CdWO$_4$ detector No. 1 over
1727 h. The fit of the time distribution gives a half-life
$T_{1/2}=(0.29\pm 0.03)~\mu$s which is in good agreement with the
table value for $^{212}$Po (0.299 $\mu$s \cite{ToI98}). }
 \label{fig:11}
\end{figure*}

Selection of double pulses produced by the fast chain of the
decays $^{212}$Bi ($ Q_\beta =2.254$ MeV) $\to $ $^{212}$Po
($Q_\alpha =8.954$ MeV, $ T_{1/2}=0$.299 $\mu $s) $\to $
$^{208}$Pb ($^{232}$Th family) was developed. An example of such
an event pulse-shape and the result of the selection for the
$^{116}$CdWO$_4$ crystal scintillator No. 1 are presented in Fig.
\ref{fig:11}. The analysis gives the activity of $^{212}$Bi (which
is in equilibrium with $^{228}$Th) 0.054(5) mBq/kg (crystal No. 1)
and 0.095(6) mBq/kg (crystal No. 2) in a reasonable agreement with
the results of the time-amplitude analysis.

\subsection{Background in the region of $Q_{\beta\beta}$ of $^{116}$Cd}

\begin{figure*}[htb]
\begin{center}
\resizebox{0.7\textwidth}{!}{\includegraphics{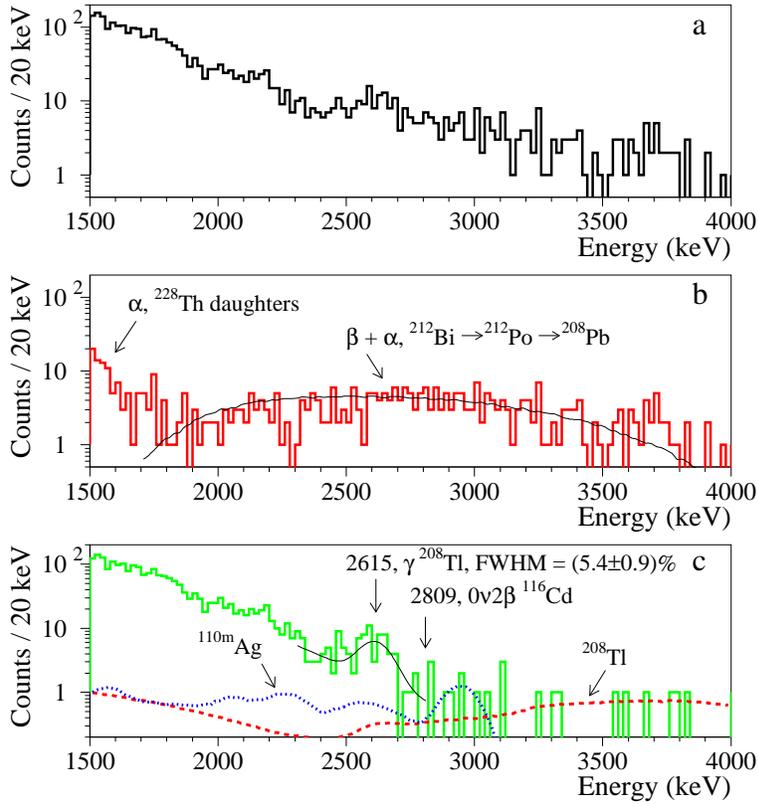}}
\end{center}
\caption{(Color online) (a) Initial sum spectrum of the two
$^{116}$CdWO$_4$ detectors measured over 1322 h (total exposure
1554 kg$\times$h) in anti-coincidence with the plastic
scintillation counter and the active light-guides; (b) the spectra
of $\alpha$ and $\beta + \alpha$ events selected by using the
pulse-shape and the front edge analyses (see text) together with
the simulated response function for the $^{212}$Bi$\to
^{212}$Po$\to ^{208}$Pb decay chain; (c) $\beta$ and $\gamma$
events selected with the help of the pulse-shape and the front
edge analyses (the efficiency of the selection procedure for
$\gamma$ quanta / $\beta$ particles is 95\%). The fit of the
$^{208}$Tl $\gamma$ peak with the energy 2615 keV is shown by
solid line. The Monte Carlo simulated energy spectra of internal
$^{110m}$Ag and $^{208}$Tl in the $^{116}$CdWO$_4$ crystals are
presented.} \label{fig:12}
\end{figure*}

By using the pulse-shape and the front edge analyses we can
substantially reduce the background of the detector in the energy
region of interest near 2.8 MeV where a peak of neutrinoless
double $\beta$ decay of $^{116}$Cd is expected (see Fig.
\ref{fig:12}). The counting rate of the detector after removing
the part of data accumulated during first 17 days of measurements
(to avoid effects of radon and cosmogenic activation) in the
energy interval of interest $2700-2900$ keV is $0.28$
counts/(yr$\times$keV$\times$kg).

A certain contribution to the background $\approx0.1$
counts/(yr$\times$keV$\times$kg) in the energy interval $2.7-2.9$
MeV comes from the 2615 keV peak caused by external $\gamma$
quanta from the decays of $^{208}$Tl (daughter of $^{232}$Th).
Contamination of the set-up (first of all of PMTs, cables, quartz
light-guides, copper shield) by thorium can be a source of the
background. We are going to simulate $^{208}$Tl decays to estimate
the contribution from different parts of the set-up. Our
preliminary calculations show that the PMTs and quartz
light-guides can be the main sources of the 2615 keV peak. We are
going to apply ultra-low background HPGe $\gamma$ ray spectrometry
to measure radioactive contamination of the PMTs and sample of the
quartz.

The two neutrino mode of $2\beta$ decay of $^{116}$Cd (assuming a
half-life $T^{2\nu}_{1/2}=2.8\times10^{19}$ yr \cite{Bar10})
contributes 0.00045 counts/(yr$\times$keV$\times$kg) in the energy
interval $2.7-2.9$ MeV.

One of the major sources of the detector background in the energy
region of the expected $0\nu2\beta$ peak of $^{116}$Cd is
contamination of the $^{116}$CdWO$_4$ crystals by $^{208}$Tl. The
Monte Carlo simulation of internal $^{208}$Tl (see Fig.
\ref{fig:12}, c) gives about $\approx0.09$
counts/(yr$\times$keV$\times$kg) in the region of interest. We
consider as a possibility the crystals recrystallization to reduce
the thorium contamination. To our knowledge there is no data in
literature about segregation of thorium to CdWO$_4$ crystals. A
positive result of recrystallization procedure can be reached if
the segregation coefficient for thorium is much less than 1. To
determine the segregation of thorium in CdWO$_4$, we are going to
measure the activity of $^{228}$Th in the $^{116}$CdWO$_4$ crystal
No. 3 by the scintillation method. Besides we are going to measure
the activity of $^{228}$Th in the rest of the melt after the
crystal growth by ultra-low background HPGe $\gamma$ ray
spectrometry, and concentration of $^{232}$Th by the High
Resolution Inductively Coupled Plasma Mass Spectrometry. Higher
concentration of thorium in the rest will be an indication of low
segregation of thorium in CdWO$_4$ crystal. Then the crystal No. 3
can be recrystallized and measured by the low counting
scintillation method to estimate activity of $^{228}$Th.

In case of positive result all the crystals can be recrystallized.
In case of not enough high efficiency of the recrystallization to
reduce thorium, one could recover the enriched cadmium from the
crystals and scraps, purify by physical (vacuum distillation and
filtering)\footnote{Unfortunately, only part of the enriched
cadmium 116 was purified by the vacuum distillation and filtering
(see subsection 2.3).} and chemical (by coprecipitation on a
collector) methods, and grow crystal again. Output of purified
materials can be decreased to reach deeper purification of
enriched cadmium. It should be stressed that a satisfactory
radiopurity level (activity of $^{228}$Th less than 0.01 mBq/kg)
was reached in some samples of CdWO$_4$ crystals
\cite{Dan96a,Bel07}. Besides, very high level of radiopurity
($\sim0.002$ mBq/kg of $^{228}$Th) was detected in zinc tungstate
(ZnWO$_4$) crystal scintillators \cite{Bel11}, which have chemical
and physical properties rather similar to CdWO$_4$. An encouraging
result of recrystallization was reported recently for calcium
tungstate (CaWO$_4$) crystal scintillators \cite{Dan11}. One could
also expect improvement of the crystal scintillators quality
thanks to deep purification of initial materials and
recrystallization. Therefore contribution from the 2615 keV peak
could be suppressed further.

It should be also mentioned a ``natural'' way of the background
decrease during next few years due to the decay of the trace
$^{228}$Th in the $^{116}$CdWO$_4$ crystals (assuming broken
equilibrium of $^{232}$Th chain and lower activity of $^{228}$Ra
in comparison to $^{228}$Th and $^{232}$Th). Such an effect
(decrease of the $^{228}$Th activity in the $^{116}$CdWO$_4$
crystals) was observed in the Solotvina experiment \cite{Dan03a}.

Finally, we assume that the essential part of the background
beyond the 2615 keV peak is due to cosmogenic activation of the
$^{116}$CdWO$_4$ crystals by $^{110m}$Ag ($Q_{\beta}=3.0$ MeV;
$T_{1/2}=250$ d) \cite{Bel01}, which can provide background up to
3 MeV. Moreover, radioactive isotope $^{110m}$Ag with activity 0.4
mBq/kg was observed in \cite{Pie94} in the enriched cadmium 116
used to produce the scintillators. Our assumption was justified by
the Monte Carlo simulation of the $^{116}$CdWO$_4$ detector
response to internal $^{110m}$Ag. A simulated model of $^{110m}$Ag
corresponding to activity 0.06 mBq/kg is shown in Fig.
\ref{fig:12}, c. Another possible cosmogenic radionuclide in the
$^{116}$CdWO$_4$ crystals can be $^{106}$Ru ($Q_{\beta}=40$ keV;
$T_{1/2}=374$ d) $\to$ $^{106}$Rh ($Q_{\beta}=3.5$ MeV;
$T_{1/2}=30$ s) \cite{Bel01}. However, cosmogenic background is
expected to be reduced substantially due to decay of cosmogenic
radionuclides, in particular of $^{110m}$Ag.

Now the experiment is in progress with an improved energy
resolution after the annealing of the crystals. We expect to
improve the background reduction at this stage of experiment
thanks to more careful pulse-shape analysis to reject events
caused by $^{228}$Th daughters. In particular, we use now a higher
resolution of the transient digitizer (50 Ms/s instead of 20 Ms/s)
to suppress further the background caused by the fast sequence
$^{212}$Bi$~\to $ $^{212}$Po $ \to $ $^{208}$Pb. In addition, we
are developing the analysis of the sequence of the decays
$^{212}$Bi $(Q_\alpha =6.207$ MeV, $T_{1/2}=60.55$ m) $\to~
^{208}$Tl $(Q_\beta =5.001$ MeV, $T_{1/2}=3.053$ m)$ \to~
^{208}$Pb to reject events of $^{208}$Tl decay. Contribution from
the 2615 keV $\gamma$ line is expected to be reduced to
$\approx0.06$ counts/(yr$\times$keV$\times$kg) thanks to the
improvement of the energy resolution.

\section{Monte Carlo simulation of double $\beta$ decay of $^{116}$Cd,
estimations of experimental sensitivity}

\begin{figure*}[htb]
\begin{center}
\resizebox{0.6\textwidth}{!}{\includegraphics{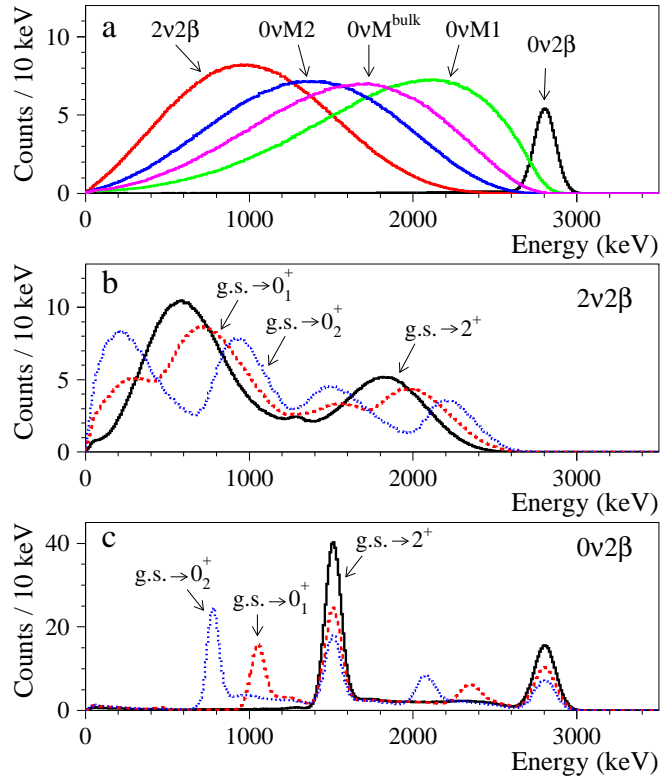}}
\end{center}
\caption{(Color online) Simulated response functions of the
$^{116}$CdWO$_4$ detector for the different modes of double
$\beta$ decay of $^{116}$Cd to the ground state of $^{116}$Sn (a);
Simulated distributions of the detector for two neutrino (b) and
neutrinoless (c) decay of $^{116}$Cd to the excited levels of
$^{116}$Sn. Approximately $3-9$ millions of decays were simulated
for the different decay channels. The distributions are normalized
to 1000 decays, except the $0\nu2\beta$ mode, which is normalized
to 100 decays.}
 \label{fig:13}
\end{figure*}

The computer simulation of the different radioactive processes in
the scintillation low background detector with $^{116}$CdWO$_4$
scintillators in the $4\pi$ active shielding has been performed
with EGS4 package \cite{EGS4}. The initial kinematics of the
particles emitted in the decay of the nuclei was given by an event
generator DECAY0 \cite{DECAY0}. The following double $\beta$
processes in $^{116}$Cd have been simulated: $0\nu2\beta$ and
$2\nu2\beta$ decay to the ground state and to the five lowest
excited levels of $^{116}$Sn; neutrinoless double $\beta$ decay
with emission of one, two and bulk \cite{Moh00} majoron(s).
Approximately $3-9$ millions of decays were simulated for the
different channels of $^{116}$Cd $2\beta$ decay. The calculated
distributions are shown in Fig. \ref{fig:13}. In particular, the
calculations give the detection efficiency in a peak of the
neutrinoless double $\beta$ decay of $^{116}$Cd as 89\% (one can
compare with the value of 83\% in the Solotvina experiment
\cite{Dan03a} where the crystals of the smaller volume were used).

On the basis of the detection efficiency and of the number of
$^{116}$Cd nuclei in the two $^{116}$CdWO$_4$ crystal
scintillators No. 1 and 2 ($1.6\times10^{24}$), assuming decrease
of the detector background to the level of $0.01-0.1$
counts/(yr$\times$keV$\times$kg) (expected due to the decay of
cosmogenic radionuclides, improvement of the energy resolution and
the pulse-shape analysis, possible change of the most contaminated
parts of the set-up and reduction of $^{228}$Th activity after
recrystallization of the $^{116}$CdWO$_4$ crystals), one can
estimate the sensitivity of a 5 years experiment to the
$0\nu2\beta$ decay of $^{116}$Cd as
$T_{1/2}\sim(0.5-1.5)\times10^{24}$ yr. According to the recent
calculations of matrix elements \cite{Kor07,Sim08,Iac11}, these
half-lives correspond to the effective neutrino mass $\langle
m_{\nu} \rangle \approx 0.4-1.4$~eV, which is on the level of the
most sensitive $2\beta$ experiments.

\section{Conclusions}

Cadmium tungstate crystal scintillators were developed from
cadmium enriched in $^{116}$Cd for a high sensitivity experiment
to search for double beta decay of $^{116}$Cd. Samples of enriched
cadmium were purified by chemical methods, the most polluted part
was additionally purified by vacuum distillation. Some part of
the enriched material was recovered from scraps after enriched
$^{116}$CdWO$_4$ crystal growth in 1988. Cadmium tungstate
compounds were synthesized from solutions. A $^{116}$CdWO$_4$
crystal boule with mass of 1868 g (which is 87\% of the initial
charge) was grown by the low-thermal-gradient Czochralski
technique. Two $^{116}$CdWO$_4$ crystal scintillators (586 g and
589 g) produced from the boule show an energy resolution
FWHM $\approx7\%$ (for 2615 keV $\gamma$ line of $^{208}$Tl) in
measurements with the crystals directly viewed by photomultiplier.

The absolute isotopic composition of $^{116}$Cd in the crystal was
determined as 82\% by mass-spectrometry. Thanks to the careful
purification of the initial materials and using the
low-thermal-gradient Czochralski technique, the crystal has rather
high transmittance: the attenuation length is $31\pm5$ cm at the
wavelength of CdWO$_4$ emission maximum 480 nm. After the low
background measurements the crystals No. 1 and 2 were annealed at
the temperature $\approx870~^{\circ}$C over 55 hours. The
annealing improved transmittance of the samples in the region of
wavelengths $350-420$ nm on $10-40~\%$.

The low background detector system with two enriched
$^{116}$CdWO$_4$ crystal scintillators (586 g and 589 g) was
installed in the underground Gran Sasso National Laboratories of
the INFN (Italy). The energy resolution of the detector with
$^{116}$CdWO$_4$ crystal scintillators inside the plastic
scintillator light-guides with length of 28 cm was even slightly
better than that in the case when the scintillators were directly
viewed by a photomultiplier. Furthermore the energy resolution was
improved to $\approx5\%$ (for 2615 keV $\gamma$ line of
$^{208}$Tl) after the annealing of the $^{116}$CdWO$_4$ crystals.

The low background measurements over 1727 h allowed to estimate
radioactive contamination of the $^{116}$CdWO$_4$ scintillators.
In addition the radioactive contamination of another
$^{116}$CdWO$_4$ sample (326 g) was tested with the help of
ultra-low background HPGe $\gamma$ ray spectrometry. The
activities of $^{226}$Ra and $^{228}$Th, which are the most
dangerous isotopes for double $\beta$ decay experiments, are on
the level $<0.005$~mBq/kg and $\approx0.06$~mBq/kg, respectively.

By using the pulse-shape discrimination of the
$^{212}$Bi$-^{212}$Po events and the anti-coinci\-de\-n\-ce signals in
the plastic scintillator light-guide and active shield, we have
obtained a background counting rate of $0.28$
counts/(yr$\times$keV$\times$kg) in the region of interest
$2700-2900$ keV.

We have tried to estimate the main sources of the detector
background on the basis of the low background measurements and the
Monte Carlo simulation: they are cosmogenic activation (most
probably $^{110m}$Ag), contamination of the $^{116}$CdWO$_{4}$
crystals by $^{228}$Th ($^{232}$Th family), contribution of the
2615 keV $\gamma$ peak of $^{208}$Tl from details of the set-up.
We are going to simulate all possible components of background;
check radioactive contamination of PMTs, quartz, cables, copper
(with aim to change the most contaminated elements); recrystallize
the crystals to reduce the concentration of thorium.

The low background measurements to search for double $\beta$ decay
of $^{116}$Cd with the help of the enriched cadmium tungstate
crystal scintillators are in progress. The decrease of thorium
concentration in $^{116}$CdWO$_4$ crystal scintillators, the decay
of cosmogenic nuclides, the improvement of the pulse-shape
analysis could further decrease the background to the level of
$\sim0.01-0.1$ counts/(yr$\times$keV$\times$kg), and therefore,
improve sensitivity of the experiment up to
$T_{1/2}\sim(0.5-1.5)\times10^{24}$ yr over 5 years of
measurements. It corresponds, taking into account the recent
calculations of matrix elements \cite{Kor07,Sim08,Iac11}, to the
effective neutrino mass $\langle m_{\nu} \rangle \approx
0.4-1.4$~eV.

Further progress in the experiment to search for the double beta
decay of $^{116}$Cd can be advanced by applying a massive array of
$^{116}$CdWO$_4$ crystals with improved energy resolution and good
particle discrimination ability.

\section{Acknowledgements}

The research was supported in part by the Gran Sasso Center for
Astroparticle Physics (L'Aquila, Italy). The group from the
Institute for Nuclear Research (Kyiv, Ukraine) was supported in
part through the Project ``Kosmomikrofizyka-2'' (Astroparticle
Physics) of the National Academy of Sciences of Ukraine. Authors
are very grateful to Prof. B.V.~Grinyov and Dr. P.N.~Zhmurin from
the Institute for Scintillation Materials (Kharkiv, Ukraine) for
kindly provided sample of LS-221 liquid scintillator. We would
like to thank Dr. C. Salvo for his kind help in preparing
equipment for the crystals annealing. It is a pleasure to
gratitude the staff of the Chemistry Service of the Gran Sasso
laboratory for their warm hospitality and valuable support.

\end{document}